\documentclass[11pt]{article}
\usepackage{geometry}                
\usepackage{graphicx}
\usepackage{amssymb}
\usepackage{amsmath}
\usepackage{amsfonts}
\usepackage{graphicx}
\usepackage{epstopdf}
\usepackage{footmisc}
\usepackage{color}
\usepackage{cite}
\usepackage{soul}

\title{Replica approach to mean-variance portfolio optimization}
\author{Istvan Varga-Haszonits$^{1}$\footnote{Present employer: MSCI, Budapest}, Fabio Caccioli$^{2,3}$ and Imre Kondor$^{1,4}$\\
{\it 1- Department of Physics of Complex Systems, E\"otv\"os University, Budapest, Hungary} \\
{\it 2- University College London, Department of Computer Science,} \\
{\it London, WC1E 6BT, UK} \\
{\it 3- Systemic Risk Centre, London School of Economics and Political Sciences, London, UK}\\
{\it 4-Parmenides Foundation, Pullach, Germany}\\
}

\begin{document}
\maketitle

\abstract{We consider the problem of mean-variance portfolio optimization for a generic covariance matrix subject to the budget constraint and the constraint for the expected return, with the application of the replica method borrowed from the statistical physics of disordered systems.  We find that the replica symmetry of the solution does not need to be assumed, but emerges as the unique solution of the optimization problem. We also check the stability of this solution and find that the eigenvalues of the Hessian are positive for $r=N/T<1$, where $N$ is the dimension of the portfolio and $T$ the length of the time series used to estimate the covariance matrix. At the critical point $r=1$ a phase transition is taking place. The out of sample estimation error blows up at this point as $1/(1-r)$, independently of the covariance matrix or the expected return, displaying  the universality not only of the critical index, but also the critical point. As a conspicuous illustration of the dangers of in-sample estimates, the optimal in-sample variance is found to vanish at the critical point inversely proportional to the divergent estimation error.}
\section{Introduction}
Starting with Markowitz's seminal paper \cite{markowitz1952Portfolio}, the problem of portfolio optimization has become the subject of a vast literature. The several hundred papers written on this problem apply widely different approaches; the list \cite{markowitz1959Portfolio,Merton1972Analytic,Dickinson1974Thereliability,jobson1979Improved,frost1986AnEmpirical,jorion1986Bayes,Konno88,konno_yamazaki,eltongruber,young1998AMinimax,artzner1999Coherent,gabor1999Portfolios,laloux1999Noise,plerou1999Universal,plerou2000ARandom,jorion,laloux2000Random,Rockafellar2000Optimization,acerbi2002Expected,acerbi2002On,pafka2002Noisy,Jagannathan2003,ledoit2003Improved,pafka2003Noisy,ledoit2004AWell,pafka2004Estimated,scherermartin,kempf2006Estimating,Okhrin2006Distributional,elton2007Modern,Garlappi2007,golosnoy2007Multivariate,Kan2007Optimal,RobustPO,kondor2007Noise,varga2007Noise,frahm2008,Basak2009Jackknife,brodie2009Sparse,DeMiguel2009,DeMiguel2009Optimal,frahm2010Dominating,kondor2010instability,still2010regularizing,Bun2016Laundrette} is just a small selection from this literature. 

The present paper belongs to a narrow subfield where the methods of statistical physics, in particular the replica method borrowed from the theory of disordered systems, are applied. In this approach one assumes that the returns on the securities in the portfolio obey an idealized probability distribution, in the simplest case a Gaussian, and calculates the quantities of interest like the weights of the optimal portfolio, the minimal risk, sensitivity to changes in the returns, estimation error, etc. exactly. Of course, real-life returns are neither normal, nor stationary, therefore one can only hope that the results produced by the theory will give an idea of the behaviour of the various characteristic quantities, which can then be compared to simulation results and empirical data. Well-suited to the case of huge institutional portfolios, the statistical physics-inspired approach targets large portfolios composed of $N$ different kind of securities, and assumes that the size $T$ of the statistical samples is comparable to $N$, so that the ratio of the two, $r=N/T$, is a fixed value. In this high-dimensional limit sample fluctuations and the concomitant estimation error can be very large, even divergent. The problem of estimation error is therefore in the focus of the statistical physics-inspired approach which, due to the highly idealized nature of the probability distribution, may be expected to provide a lower bound for this error. 

The first step along this path was taken by Ciliberti et al \cite{Ciliberti2007On} who studied the large fluctuations and the resulting phase transition in the case of the special risk measure Expected Shortfall (ES), soon to be followed by a similar work \cite{ciliberti2007Risk} on the mean absolute deviation. In subsequent papers ES was optimized under various regularizers to suppress the instability of estimation, and to take into account the future market impact of an eventual liquidation of the portfolio \cite{caccioli2013Optimal}, \cite{caccioli2016Lp}. The works \cite{Ciliberti2007On}, \cite{ciliberti2007Risk}, \cite{caccioli2013Optimal}, \cite{caccioli2016Lp} were mostly concerned with the instability and the accompanying phase transition, accordingly they considered i.i.d. returns (i.e. a diagonal covariance matrix with identical elements), so that to show up the phenomenon in the simplest possible setting. The papers \cite{kondor2015Contour}, \cite{caccioli2015Portfolio} changed focus, and still assuming i.i.d. returns, they mapped out the whole parameter space of ES, demonstrating the very serious estimation error problem even far away from the instability region. Finally, \cite{Papp2016Variance} considered also an $\ell_2$ regularized version of ES, which allowed an insight into how variance-bias tradeoff is playing out in the case of this particular risk measure. 

The prominence of the risk measure ES in all these papers was mainly motivated by its anticipated regulatory significance; and indeed, after a long consultation period the Basel Committee finally instituted ES as the global regulatory market risk measure in January 2016 \cite{basle2016Minimum}. Because of the focus on the regulatory context, the application of the replica method to the most obvious and simplest risk measure, the variance, has been left in the background. Nevertheless, it was not completely neglected: we performed the optimization of variance via replicas quite some years ago, but the results were left unpublished.\footnote{The results for variance formed part of the PhD thesis  ``The instability of risk measures'' (in Hungarian) by one of us (I.V-H.), written under the supervision of I.K. and successfully defended at E\"otv\"os University, Budapest, in 2009. Independently, F.C. also performed the replica treatment of the variance optimization at the Santa Fe Institute in 2011.} As we recently decided to carry out a systematic study of the sensitivity of different risk measures to estimation error (as it were, an analytical counterpart of the numerical study \cite{kondor2007Noise}) and also their responses to various regularization schemes, it is high time to publish the results for the variance now.\footnote{During the preparation of this manuscript we came to learn about T. Shinzato's preprint posted on arXive in May 2016 \cite{Shinzato2016Minimal}, which is concerned with the replica optimization of the variance at the global minimum in the special case of i.i.d. normal returns and with a constraint promoting the concentration of investment, equivalent to an $\ell_2$ regularizer. Despite the similar subject and method, there is little overlap between his work and ours.}

While the papers quoted in the previous paragraph all confined their treatment to the simplified case of independent random variables (diagonal covariance matrix) and studied the global minimal portfolio, here we present the replica treatment of the full Markowitz problem, that is we consider a generic full rank covariance matrix, and take into account not only the budget constraint but also the constraint on the expected return. Thereby we calculate the estimation error all along the efficient frontier. 

The optimization of variance leads us to two very interesting results. In contrast to the usual situation, where at a certain point in the derivation one has to choose the replica symmetric minimum from among several seemingly possible ones and hope this choice leads to the correct solution, in the special case of the variance the symmetric solution emerges as the unique minimum. Since the variance is convex, it necessarily has a unique minimum, therefore the unique replica symmetric minimum found by the method must coincide with the correct result. The other remarkable feature of the solution is that the estimation error turns out to be the same along the whole efficient boundary, independently of the covariance matrix and the expected return.\footnote{Various special cases of this invariance have long been known to us from earlier numerical \cite{pafka2004Estimated} and analytical work (B. Komjati: The instability of portfolio selection (in Hungarian), MSc thesis, E\"otv\"os University, Budapest, 2008), but the result to be presented below is the most general proof of this surprising invariance.} In principle, one might have expected that a covariance matrix with several more or less strongly correlated returns (but still full rank) would display a reduced effective dimension, hence a shift in the critical point. The independence of the estimation error from the covariances and expected returns is the manifestation of a surprising degree of universality, analogous to the independence of the critical point of the minimax risk measure from the underlying probability distribution (see \cite{kondor2007Noise}), but also to the universality discovered in a wide class of high dimensional random geometric phase transitions by Donoho and Tanner \cite{Donoho2009Observed}. 

When looking for a minimum, one should, in general, check the behaviour of the second derivatives. Although there could be no doubt about the extremum being a stable minimum in the case of the variance, we nevertheless calculate the Hessian (the matrix of second derivatives) at the extremum, and show that all its eigenvalues are positive. This is meant to be an example for a check one should always perform, but often neglects. 

Our calculations are limited to the case where both the number $N$ of assets in the portfolio and the sample size $T$ (the length of the time series for the returns) are large, with their ratio $r=N/T$ a fixed number, less than 1. ($r=1$ is the critical point of the problem, beyond which the covariance matrix ceases to be of full rank, and the optimization problem becomes meaningless.)  We note that in the case of the variance, finite $N$ and finite $T$ results also exist (e.g. \cite{Okhrin2006Distributional}, \cite{Basak2009Jackknife}, \cite{frahm2010Dominating}) and in the appropriate limit they coincide with our results. 

We also note that the Markowitz problem with the expected return constraint relaxed is equivalent to the problem of linear regression, see e.g. \cite{kempf2006Estimating}, \cite{kondor2008Divergent}, thus the replica approach can naturally cover regression type problems as well.

The plan of the paper is as follows. In Sec. 2 we formulate the statistical mechanics of portfolio optimization assuming that the covariance matrix and returns are known. Sec. 3 is the replica treatment of the problem when the returns are drawn from a multivariate normal distribution, with a generic covariance matrix, a budget constraint and a constraint on the expected return. Finally, Sec. 4 is a short summary. As the main purpose of the paper is to exhibit the technique of replicas in the context of variance optimization, we include some of the details of the calculation in the main text rather than exiling them to Appendices.

\section{The Markowitz problem as a statistical physics model}
\label{sec:pfstatphys}

Portfolio optimization is a tradeoff between risk and return: it seeks to maximize the return of a portfolio at a given level of risk, or minimize the risk at a given level of return. If, following Markowitz [1], we assume that the returns are drawn from a multivariate Gaussian distribution and risk is measured in terms of the variance of the portfolio, then we are led to the following quadratic optimization problem: 
\begin{align}
\min_{\mathbf{w}\in\mathbb{R}^N}&\sum_{i=1}^N\sum_{j=1}^N\sigma_{ij}w_iw_j,\label{eq:markowitzmin}\\
&\sum_{i=1}^Nw_i\mu_i=\mu,\label{eq:markowitzreturn}\\
&\sum_{i=1}^Nw_i=1,\label{eq:markowitzbudget}
\end{align}
where $N$ is the number of securities in the portfolio, $\mu_i$ is the expected value of the return on security $i$, $\mu$ is the expected return on the portfolio, $\sigma_{ij}$ is the covariance matrix of returns, and $w_i$ is the weight of security $i$ in the portfolio. The solution of this problem is the set of weights that minimize the variance given a fixed return $\mu$ and a fixed budget expressed by the sum of weights being fixed at 1. Note that in this simple setup the weights are not constrained to be positive, so unlimited short selling is allowed. 

We will refer to the above optimization problem as the Markowitz problem. If the covariance matrix and the expected returns are given, the Markowitz problem can be easily solved by the method of Lagrange multiplyers. As the covariance matrix is positive definite, the objective function is strictly convex, so the problem has a unique solution. This solution was first derived by Merton \cite{Merton1972Analytic}.

Introducing the notations
$A=\sum_{i,j}\sigma^{-1}_{ij}$, $B=\sum_{i,j}\sigma^{-1}_{ij}\mu_j$ and $C=\sum_{i,j}\sigma^{-1}_{ij}\mu_i\mu_j$ the solution is:
\begin{align}
w^*_i(\mu)&=\sum_{j=1}^N\sigma^{-1}_{ij}\left[\lambda^{*}(\mu)+\eta^{*}(\mu)\mu_j\right],\label{eq:optpf}\\
\lambda^{*}(\mu)&=\frac{C-B\mu}{AC-B^2},\label{eq:optlambda}\\
\eta^{*}(\mu)&=\frac{A\mu-B}{AC-B^2}.\label{eq:opteta}
\end{align}
The variance of the optimal portfolio for a given expected return $\mu$ is:
\begin{equation}
{\sigma^*}^2(\mu)=\sum_{i=1}^N\sum_{j=1}^N\sigma_{ij}w^*_i(\mu)w^*_j(\mu)=\frac{A}{AC-B^2}\left(\mu-\frac{B}{A}\right)^2+\frac{1}{A},\label{eq:frontier}
\end{equation}
where $AC-B^2>0$.

\begin{figure}
\begin{center}
\includegraphics[width=10cm]{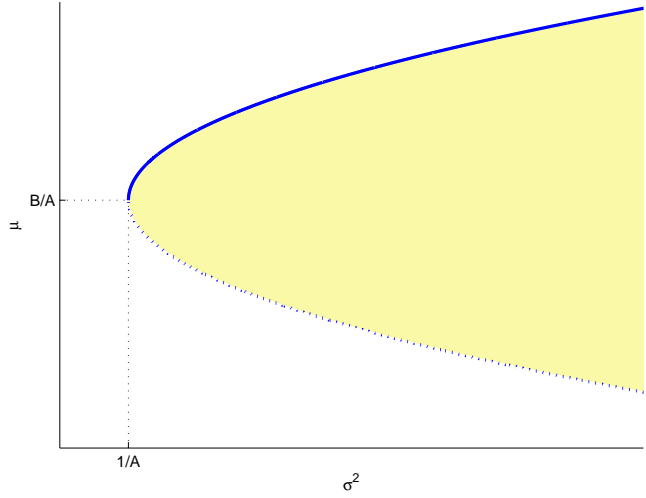}
\caption{\label{fig:frontier} The set of attainable portfolios (yellow area), marginal portfolios (continuous and dotted blue line) and efficient portfolios (continuous blue line) on the risk-return plane.}
\end{center}
\end{figure}
 
The financial meaning of the result is illustrated in Figure \ref{fig:frontier}. The set of those portfolios whose expected return $\mu$ and risk $\sigma$ satisfy the inequality $\sigma^2\ge{\sigma^*}^2(\mu)$ are called attainable portfolios, because these are the ones that can be composed out of the $N$ securities under the budget constraint \eqref{eq:markowitzbudget}. The solutions of the constrained optimization problem \eqref{eq:markowitzmin} are called marginal portfolios, because for these $\sigma^2={\sigma^*}^2(\mu)$ holds, so they constitute the boundary of the set of attainable portfolios.

As shown by the figure, if $\sigma^2>1/A$ there are two marginal portfolios for every value of the standard deviation $\sigma$. Of these, the one with the larger expected return will be efficient, since for a given risk this will have the largest return among the attainable portfolios. The efficient portfolios will therefore be those for which the conditions $\mu>B/A$ and $\sigma^2={\sigma^*}^2(\mu)$ are fulfilled simultaneously. These are shown by the continuous blue line in the figure.

In order to provide a bridge to the formalism presented in the next section, we now rephrase the standard Markowitz problem of optimal portfolio selection. Following Kirkpatrick et al. \cite{Kirkpatrick1983Simulated}, we convert the problem of optimizing the variance into a problem in statistical physics. For this purpose, we regard the variance (with a factor $\frac{1}{2}$ included for convenience) as the Hamiltonian of a fictitious physical system:
\begin{equation}
\mathcal{H}=\frac{1}{2}\sum_{i=1}^N\sum_{j=1}^N\sigma_{ij}w_iw_j+\sum_{i=1}^Nh_iw_i,
\end{equation} 
where $w_i$ are the portfolio weights and $\sigma_{ij}$ are the elements of the covariance matrix, which, in this Section, we continue to regard as given. The $h_i$'s are external fields conjugate to the variables $w_i$. In the case of $h_1=h_2=...=h_N=0$ the Hamiltonian is half of the risk (measured in terms of the variance) of the portfolio $\mathbf{w}$. The optimization of the variance corresponds to finding the ground state of the above Hamiltonian. 

Assume the system is subject to a Boltzmann distribution at inverse temperature $\beta$. Imposing the budget constraint $\sum_iw_i=1$ and the constraint on the expected return $\sum_i\mu_iw_i=\mu$ the partition function can be written as:
\begin{equation}
Z=\int_{-\infty}^{\infty}\prod_i dw_i \exp\left(-\frac{1}{2}\beta\sum_{i,j}\sigma_{ij}w_iw_j-\beta\sum_ih_iw_i\right)
\delta\left(\sum_iw_i-1\right)\delta\left(\sum_i\mu_iw_i-\mu\right),\label{eq:partfunc}
\end{equation}
where the index $i$ runs from $1$ to $N$. By the Fourier representation of the  Dirac-$\delta$ we obtain the following Gaussian integral:
\begin{multline}
Z=\int_{-\infty}^{\infty}d\eta\int_{-\infty}^{\infty}d\lambda \int_{-\infty}^{\infty}\prod_i dw_i \exp\left(-\frac{1}{2}\beta\sum_{i,j}\sigma_{ij}w_iw_j-\beta\sum_ih_iw_i\right)\times\\
\times\exp\left[-i\lambda\left(\sum_iw_i-1\right)-i\eta\left(\sum_i\mu_iw_i-\mu\right)\right].
\end{multline}
Having calculated the integral, the free energy, defined by the relation $F=-\beta^{-1}\ln Z$, can be written as:
\begin{multline}
F=-\frac{N}{2\beta}\ln\left(\frac{2\pi}{\beta}\right)+\frac{1}{2\beta}\mathrm{Tr}\ln\left(\sigma\right)-\frac{1}{2\beta}\ln\left(AC-B^2\right)-\frac{1}{2}\sum_{i=1}^N\sum_{j=1}^N\sigma^{-1}_{ij}h_ih_j+\\
+\frac{1}{AC-B^2}\left[A\left(\sum_{i,j}\sigma^{-1}_{ij}\mu_ih_j+\mu\right)^2+C\left(\sum_{i,j}\sigma^{-1}_{ij}h_j+1\right)^2-\right.\\
\left.-2B\left(\sum_{i,j}\sigma^{-1}_{ij}\mu_ih_j+\mu\right)\left(\sum_{i,j}\sigma^{-1}_{ij}h_j+1\right)\right],\label{eq:helmholtz}
\end{multline}
where we used the notations $A=\sum_{i,j}\sigma^{-1}_{ij}$, $B=\sum_{i,j}\sigma^{-1}_{ij}\mu_j$ and $C=\sum_{i,j}\sigma^{-1}_{ij}\mu_i\mu_j$ . The weights of the optimal portfolio are the ensemble averages of the variables $w_i$ at zero temperature and zero external field. They can be obtained by taking the derivative of the free energy with respect to the conjugate fields $h_i$:
\begin{equation}
\left.\frac{\partial F}{\partial h_i}\right|_{h_1=h_2=...=h_N=0}=\langle w_i\rangle,
\end{equation}
which gives
\begin{equation}
\langle w_i\rangle=\sum_{j}\sigma^{-1}_{ij}\left(\frac{A\mu-B}{AC-B^2}\mu_j+\frac{C-B\mu}{AC-B^2}\right).
\end{equation}
It is remarkable that $\langle w_i\rangle$ does not depend on $\beta$, thus the thermal average produces the optimal weights not only at zero, but also at finite temperature. (A side remark: this is like the thermal average of the atoms' positions in a harmonic solid. Thermal expansion needs anharmonic terms.) As we shall see, however, the curve of the efficient portfolios can be recovered only in the low temperature limit. For this, let us write the free energy in zero field:
\begin{equation}
F=-\frac{N}{2\beta}\ln\left(\frac{2\pi}{\beta}\right)+\frac{1}{2\beta}\mathrm{Tr}\ln\left(\sigma\right)-\frac{1}{2\beta}\ln\left(AC-B^2\right)+\frac{1}{2}\frac{1}{AC-B^2}\left(A\mu^2-2B\mu+C\right).
\end{equation}
At zero temperature the free energy is the minimum of the Hamiltonian, that is the optimal value of the risk given the expected value of the return $\mu$ :
\begin{equation}
\sigma^{*}(\mu)=2\lim_{\beta\to\infty}F=\frac{A\mu^2-2B\mu+C}{AC-B^2},
\end{equation}
which, after some rearrangement, is seen to agree with the formula for the efficient frontier derived above.

\section{Sensitivity to noise and the replica method}

In contrast to the assumption in the previous section, in real life the returns and covariances are not known, but have to be estimated from empirical samples. Assume, therefore, that we observe the price changes of our securities in $T$ ($T<\infty$) subsequent (nonoverlapping and equal) periods, and denote the relative price change of security $i$ in period $t$ by $x_{it}$  ($i=1,2,...,N$ és $t=1,2,...,T$). Of course, the returns $x_{it}$ will also be normally distributed variables, furthermore $\mathbb{E}[x_{it}]=\mu_i$ (for all $t$) and $\mathbb{E}[x_{it}x_{js}]-\mathbb{E}[x_{it}]\mathbb{E}[x_{js}]=\sigma_{ij}\delta_{ts}$, because the returns are assumed serially independent (their autocorrelation is zero). Accordingly, the unbiased estimate of the parameters of the distribution is given by the formulae
\begin{align}
&\hat{\mu}_i=\frac{1}{T}\sum_{t=1}^Tx_{it},\label{eq:samplemean}\\
&\hat{\sigma}_{ij}=\frac{1}{T-1}\sum_{t=1}^{T}\left(x_{it}-\hat{\mu}_i\right)\left(x_{jt}-\hat{\mu}_j\right)\label{eq:samplecovmat}
\end{align}
Replacing the true distribution by the estimated one based on a finite sample will unavoidably introduce estimation error in the values of the optimal weights and also in the value of the variance itself. We will denote the optimal estimated weights by $\hat{w}^*_i$ and the estimated optimal variance by $(\hat{\sigma}^*)^2$ in the following. Pafka and Kondor \cite{pafka2003Noisy} (see also the considerations in \cite{caccioli2015Portfolio}) introduced the quantity
\begin{equation}
q_0=\frac{\sum_{i,j}\sigma_{ij}\hat{w}^*_i\hat{w}^*_j}{\sum_{i,j}\sigma_{ij}w^*_iw^*_j}.\label{eq:q0std}
\end{equation}
as a measure of out of sample estimation error. As in the numerator we have the true covariance matrix multiplied by the estimated weights, whereas in the denominator the same true covariance matrix is multiplied by the ``true'' weights (that minimize the variance), it is clear that $q_0\ge1$, where for finite samples the equality holds with zero probability. Thus the number $\sqrt{q_0}-1$ determines the relative error in the risk in the estimated portfolio. Since the estimated covariance matrix and the corresponding optimal weights fluctuate from sample to sample, the quantity $q_0$ itself will also be a random variable. The first two moments of $q_0$ can give an idea about how sensitive the Markowitz model is to estimation error. One of the main results of the present calculation will be to derive the expectation value of $q_0$ for an arbitrary covariance matrix and verify its universality (its independence of the covariance matrix and the return), a possibility first raised on the basis of numerical experiments by \cite{pafka2004Estimated}. As for the second moment, its behaviour can be inferred from the results in \cite{Okhrin2006Distributional}, \cite{Basak2009Jackknife}, or  \cite{frahm2010Dominating}: in the thermodynamic limit the variance of $q_0$ vanishes, in the parlance of the theory of disordered systems, $q_0$ self-averages.

In addition to the estimation error, we may even worry about whether the optimization can be carried out at all, that is whether the estimated covariance matrix $\hat{\sigma}_{ij}$ preserves the positive definitness of the true $\sigma_{ij}$, so that to remain invertible. According to elementary linear algebra, the condition for the positive definiteness of the estimated covariance matrix is $T\ge N$, because the rank of $\hat{\sigma}_{ij}$ is, with probability one, equal to $\min\{N,T\}$. This means we must have at least as many, or more, observations for each security as the dimension $N$ of the portfolio, a very natural requirement. Accordingly, the ratio $r=N/T$ will play a crucial role in the following. For very small values of $r$, when we have plenty of data, we are in the realm of ordinary statistics, and by force of the central limit theorem our estimates will converge to their true values. If $r$ is not small enough, we will be working in the high dimensional regime, where the estimates may strongly deviate from the true values, and as we approach $r=1$ from below we may expect the estimation error to blow up. Beyond this critical value of $r$ the optimization of the variance cannot be performed. During the long history of portfolio theory, financial mathematics, statistics and computer science introduced a plethora of methods to deal with this difficulty; ultimately all these procedures boil down to a modification of the original problem (via dimensional reduction, regularization, etc.) so as to remove the instability at the price of permitting a, hopefully limited, bias. As the purpose of the present paper is to demonstrate the application of statistical physics methods to the optimization of the variance, we do not concern ourselves with these methods here, and keep to the original framework of mean-variance optimization.

A characteristic feature of the replica method that we are going to apply in the following is that at a certain point one has to calculate a saddle point integral where the dimension $N$ is let go to infinity. Then, to keep the ratio $r$ below its critical value 1, we have to consider very large samples. This means we will be interested in the ``thermodynamic'' limit where both $N$ and $T$ are large, but their ratio stays finite, smaller than 1. For this reason, we may safely neglect the 1 in the denominator in \eqref{eq:samplecovmat}.

Let us now consider the optimization of the variance estimated from empirical samples. Introducing the notation $u_t=\sum_iw_i(x_{it}-\mu_i)$ the problem can be formulated as
\begin{align}
\min_{\mathbf{u},\mathbf{w}}\frac{1}{T}&\sum_{t=1}^T\left(u_t-\frac{1}{T}\sum_{s=1}^Tu_s\right)^2,\\
&\sum_{i=1}^Nw_i(x_{it}-\mu_i)=u_t,\\
\frac{1}{T}&\sum_{t=1}^T\sum_{i=1}^Nw_ix_{it}=N\mu,\\
&\sum_{i=1}^Nw_i=N.\label{eq:pfnorm}
\end{align}
Note the change of normalization in the budget constraint \eqref{eq:pfnorm}. The purpose of this modification is to ensure that the order of magnitude of the weights remain $O(1)$ in the thermodynamic limit. Let us write up the partition function corresponding to this problem, following the recipe in the previous section:
\begin{multline}
Z \propto  \int\prod_tdu_t \int\prod_td\hat{u}_t e^{-\frac{\beta}{2T}\sum_t\left(u_t-\frac{1}{T}\sum_su_s\right)^2}\int\prod_idw_i e^{i\sum_t\hat{u}_t\left[u_t-\sum_iw_i(x_{it}-\mu_i)\right]}\\
\int d\eta e^{-i\eta\left(\frac{1}{T}\sum_tu_t+\sum_iw_i\mu_i-N\mu\right)}\int d\lambda e^{-i\lambda\left(\sum_iw_i-N\right)},\label{eq:partfuncnoise}
\end{multline}
where the integrals run from $-\infty$ to $\infty$. We have omitted here an unimportant constant factor in $Z$ (hence the notation $\propto$ instead of $=$). We will do so also in the following.

We wish to average $Z$ over the random sample $\{x_{it}\}$ whose joint distribution function is:
\begin{equation}
f\left(\{x_{it}\}\right)=(2\pi)^{-NT/2}\left(\det\sigma\right)^{-T/2}e^{-\frac{1}{2}\sum_{i,j,t}\sigma_{ij}^{-1}(x_{it}-\mu_i)(x_{jt}-\mu_j)}.\label{eq:gausspdf}
\end{equation}
We assume that $\sigma_{ij}$ is strictly positive definite. In order to facilitate the derivation, we express the variables $x_{ij}$ through the standard normal variables $z_{it}$:
\begin{equation}
x_{it}=\sum_jD_{ij}z_{jt}+\mu_i,\label{eq:standardize}
\end{equation}
where $D_{ij}$ is the Cholesky decomposition of the true covariance matrix $\sigma_{ij}$ of the returns, that is, by definition  $\sigma_{ij}=\sum_{k}D_{ik}D_{jk}$. On the basis of the formula \eqref{eq:gausspdf} we can see, by a simple change of variables, that the joint probability distribution of the random variables $z_{it}$ is indeed
\begin{equation}
f\left(\{z_{it}\}\right)=(2\pi)^{-NT/2}e^{-\frac{1}{2}\sum_{i,t}z_{it}^2}.\label{eq:stdgausspdf}
\end{equation}
Expressed through the standard normal variables $z_{it}$ the partition function becomes 
\begin{multline}
Z \propto  \int\prod_tdu_t \int\prod_td\hat{u}_t e^{-\frac{\beta}{2T}\sum_t\left(u_t-\frac{1}{T}\sum_su_s\right)^2}\int\prod_idv_i e^{i\sum_t\hat{u}_t\left(u_t-\sum_iv_iz_{it}\right)}\\
\int d\eta e^{-i\eta\left(\frac{1}{T}\sum_tu_t+\sum_iv_i\theta_i-N\mu\right)}\int d\lambda e^{-i\lambda\left(\sum_iv_id_i-N\right)},\label{eq:stdpartfunc}
\end{multline}
where we have changed variables $v_i=\sum_jw_jD_{ji}$ , and introduced the notations $d_i=\sum_jD^{-1}_{ij}$ and $\theta_i=\sum_jD^{-1}_{ij}\mu_j$ .

\subsection{The replica method}
\label{sec:replicatrick}

Our goal is to calculate the average of free energy density over the samples in the thermodynamic limit ($N/T$ constant and $N\to\infty$):
\begin{equation}
\mathbb{E}[f]=-\lim_{N\to\infty}\frac{1}{N\beta}\mathbb{E}\left[\ln Z\right].
\end{equation}
The direct calculation of the expected value of the logarithm of a random variable is difficult, but by the relation
\begin{equation}
\ln Z=\lim_{n\to0}\frac{Z^n-1}{n}\label{eq:replicatrick}
\end{equation}
we can reduce the problem to the calculation of $\mathbb{E}[Z^n]$. If $n$ is a natural number then $Z^n$ is the partition function of $n$ identical, independent copies or replicas (hence the name of the method) of the original system, which by \eqref{eq:stdpartfunc} can be written as:
\begin{multline}
Z^n \propto  \int\prod_{a,t}du_t^a \int\prod_{a,t}d\hat{u}_t^a e^{-\frac{\beta}{2T}\sum_{a,t}\left(u_t^a-\frac{1}{T}\sum_su_s^a\right)^2}\int\prod_{a,i}dv_i^a e^{i\sum_{a,t}\hat{u}_t^a\left(u_t^a-\sum_iv_i^az_{it}\right)}\\
\int\prod_a d\eta^a e^{-i\sum_a\eta^a\left(\frac{1}{T}\sum_tu_t^a+\sum_iv_i^a\theta_i-N\mu\right)}\int\prod_a d\lambda^a e^{-i\sum_a\lambda^a\left(\sum_iv_i^ad_i-N\right)},\label{eq:replicapartfunc}
\end{multline}
The replica indices $a$ run from $1$ to $n$. This expression can be easily averaged over the the $z_{it}$. In the following, after performing the integrals, we shall bring the replica-partition function to the form $\mathbb{E}[Z^n]\propto\exp\left(-\beta Nng(\beta,N/T,N)+O(n^2)\right)$ . Substituting this expression into the equation \eqref{eq:replicatrick}, and reinterpreting the number $n$ as a real, we can carry out the limit $n\to0$. As a result, we find that the function $g$ in the exponent is, up to an unimportant additive constant, the average of the free energy density, that is $\mathbb{E}[f(\beta,r)]=const+\lim_{N\to\infty}g(\beta,r,N)$, where we used the notation $r=N/T$.

The meaning of the expression ``reinterpreting the number $n$ as a real'' is that we perform an analytical continuation from the set of natural numbers to the reals, an operation whose result is not necessarily unique. The analytic continuation is the weak link in the chain of manipulations making up the replica method; without a rigorous proof of the uniqueness of analytic continuation the method can only be regarded as heuristic. A guarantee of the uniqueness of the continuation can only come from imposing some additional constraints on the problem. We conjecture that this constraint is the convexity of the objective function, which guarantees that the optimization problem has a single solution, and it is hard to imagine how the analytic continuation could lead to a different one. As the variance is convex, the replica method should produce the correct results for its minimization. This, of course, does not constitute a proof, but rigorous proofs exist in similar, and even much more complicated, problems in the theory of disordered systems \cite{guerra2002TheThermodynamic}, \cite{guerra2003TheInfinite}, \cite{talagrand2003spin}, which lends support to our conjecture. Furthermore, the results to be derived below agree with the rigorous results by \cite{Okhrin2006Distributional}, \cite{Basak2009Jackknife}, \cite{frahm2010Dominating} in the thermodynamic limit, and also with the numerical experiments by \cite{kondor2007Noise}.

\subsection{Averaging over the samples}
\label{sec:quenchedavg}

Let us average the replica partition function \eqref{eq:replicapartfunc} with the density function \eqref{eq:stdgausspdf}. For this, we have to calculate the following integral:
\begin{equation}
(2\pi)^{-NT/2}\int\prod_{i,t}dz_{it}e^{-\frac{1}{2}\sum_{i,t}z_{it}^2-i\sum_{a,i,t}\hat{u}_t^av_i^az_{it}}=e^{-\frac{1}{2}\sum_{it}\left(\sum_av_i^a\hat{u}_t^a\right)^2}.\label{eq:quavg}
\end{equation}
Let us introduce the overlap matrix by:
\begin{equation}
Q^{ab}=\frac{1}{N}\sum_{i=1}^Nv_i^av_i^b,
\end{equation}
and transform \eqref{eq:quavg} as follows:
\begin{multline}
e^{-\frac{1}{2N}\sum_{it}\left(\sum_av_i^a\hat{u}_t^a\right)^2}\propto\int\prod_{ab}dQ^{ab}\int\prod_{ab}d\tilde{Q}^{ab}\\
e^{\frac{i}{2}N\sum_{a,b}\tilde{Q}^{ab}\left(Q^{ab}-\frac{1}{N}\sum_iv_i^av_i^b\right)-\frac{N}{2}\sum_{ab}Q^{ab}\sum_t\hat{u}_t^a\hat{u}_t^b}.
\end{multline}
Plugging this back into \eqref{eq:replicapartfunc}, after some rearrangement we obtain
\begin{multline}
\mathbb{E}[Z^n] \propto \int\prod_{ab}dQ^{ab}\int\prod_{ab}d\hat{Q}^{ab}e^{\frac{1}{2}N\sum_{ab}\hat{Q}^{ab}Q^{ab}}
\int\prod_{a,t}du_t^a e^{-\frac{\beta}{2T}\sum_{a,t}\left(u_t^a-\frac{1}{T}\sum_su_s^a\right)^2}\\ \int\prod_{a,t}d\hat{u}_t^a e^{-\frac{N}{2}\sum_{a,b}Q^{ab}\sum_t\hat{u}_t^a\hat{u}_t^b+i\sum_{a,t}\hat{u}_t^a u_t^a}
\int\prod_a d\eta^a \int\prod_a d\lambda^a e^{i\sum_a\eta^a\left(N\mu-\frac{1}{T}\sum_tu_t^a\right)+iN\sum_a\lambda^a}\\
\int\prod_{a,i}dv_i^a e^{-\frac{1}{2}\sum_{a,b}\hat{Q}^{ab}\sum_iv_i^av_i^b-i\sum_{a,i}\left(\eta^a\theta_i+\lambda^ad_i\right)v_i^a},\label{eq:avgpartfunc}
\end{multline}
where we have applied the replacement $\hat{Q}^{ab}=i\tilde{Q}^{ab}$ ; accordingly the integration with respect to the variables $\hat{Q}^{ab}$ is along the imaginary axis from $-i\infty$ to $i\infty$. The Gaussian integrals over $\hat{u}_t^a$, $v_i^a$, $\eta^a$ and $\lambda^a$ can then be performed easily. The only thing to watch out for is that as $\hat{Q}^{ab}$ is imaginary, in some of the Gaussian integrals the parameter standing in the place of the standard deviation will also be imaginary. This, however, does not pose a problem, because the contour of integration over $\hat{Q}^{ab}$ can be deformed so as to make it run to the right of the imaginary axis and return to the imaginary axis only at $\pm i \infty$. Thus, for finite values of $\hat{Q}^{ab}$  $\mathrm{Re}\left(\hat{Q}^{ab}\right)>0$ and all the Gaussian integrals will be meaningful. Finally, we end up with the result:
\begin{equation}
\mathbb{E}[Z^n]\propto \int\prod_{ab}dQ^{ab}\int\prod_{ab}d\hat{Q}^{ab}e^{-N\left[G(\mathbf{Q},\hat{\mathbf{Q}};\beta)+O(1/N)\right]}, \label{eq:beforesp}\\
\end{equation}
where
\begin{multline}
G(\mathbf{Q},\hat{\mathbf{Q}};\beta)=\lim_{N\to\infty}\left[-\frac{1}{2}\sum_{a,b}\hat{Q}^{ab}\left(Q^{ab}-N{\sigma^*}^2(\mu)\right)+\right.\\
\left.+\frac{1}{2r}\mathrm{Tr}\ln\mathbf{Q}+\frac{1}{2}\mathrm{Tr}\ln\hat{\mathbf{Q}}-\frac{1}{N}\ln A(\mathbf{Q},\hat{\mathbf{Q}};\beta)\right],\label{eq:replicaexp}
\end{multline}
\begin{multline}
A(\mathbf{Q},\hat{\mathbf{Q}};\beta)=\int\prod_{a,t}du_a^te^{-\frac{\beta}{2T}\sum_{at}\left(u_t^a-\frac{1}{T}\sum_su_s^a\right)^2-\frac{1}{2N}\sum_{a,b}\left[\mathbf{Q}^{-1}\right]^{ab}\sum_tu_t^au_t^b}\\
e^{-\frac{1}{2}\sum_{ab}\hat{Q}^{ab}\left[\frac{\alpha^*}{T^2}\sum_{t,s}u_t^au_s^b-\frac{N\eta^{*}(\mu)}{T}\sum_t(u_t^a+u_t^b)\right]}.
\end{multline}
and the limit $N\to\infty$ is taken such that $r=N/T=const$. The quantities $\eta^*(\mu)$  and $\sigma^{*}(\mu)$ are defined by \eqref{eq:opteta},  and \eqref{eq:frontier}, respectively, and, in terms of the notations in Sec. 2,  $\alpha^*=A/(AC-B^2)$.  As a reminder: $N\sigma^{*}(\mu)$ is nothing but the true risk of the portfolio (corresponding to $r\to 0$). (The factor $N$ appears because of the modified budget constraint). In the following, we shall assume that ${\sigma^{*}}^2(\mu)$, $\eta^*(\mu)$ and $\alpha^*$ are of the order of $O(1/N)$. (For example, for $\sigma_{ij}=\delta_{ij}$ and $\mu_i=const$ this is so automatically.) It can be shown that this is merely a technical assumption which does not influence the validity of the results.

As we want to determine $A(\mathbf{Q},\hat{\mathbf{Q}};\beta)$ only in the thermodynamic limit we can omit the terms less than $O(N)$ to obtain:
\begin{multline}
A(\mathbf{Q},\hat{\mathbf{Q}};\beta)=\int\prod_{a,t}du_a^te^{-\frac{1}{2N}\sum_{a,b,t,s}\left[\left(\beta r+[\mathbf{Q}^{-1}]^{ab}\right)\delta_{ts}+\frac{r^2}{N}\left(\alpha^*\hat{Q}^{ab}-\beta\delta^{ab}\right)\right]u_t^au_s^b+\eta^*(\mu)r\sum_{a,b,t}\hat{Q}^{ab}u_t^b}\propto\\
\propto e^{-\frac{N}{2r}\left[\mathrm{Tr}\ln\left(\beta r\mathbf{I}+\mathbf{Q}^{-1}\right)+O(1/N)\right]},
\end{multline}
where $\mathbf{I}$ is the $n\times n$ identity matrix. Plugging this result back into \eqref{eq:replicaexp} we get:
\begin{equation}
G(\mathbf{Q},\hat{\mathbf{Q}};\beta)=\frac{1}{2}\left[-\sum_{a,b}\hat{Q}^{ab}\left(Q^{ab}-\nu(\mu)\right)+\mathrm{Tr}\ln\hat{\mathbf{Q}}+\frac{1}{r}\mathrm{Tr}\ln\left(\beta r\mathbf{Q}+\mathbf{I}\right)\right],\label{eq:exponent}
\end{equation}
where $\nu(\mu)=\lim_{N\to\infty}N{\sigma^*}^2(\mu)$.

\subsection{The ``physical'' meaning of the overlap matrix}

Before continuing, let us take a closer look at how the matrix elements of $\mathbf{Q}$ can be interpreted. Let us consider two replicas with indices $a$ and $b$. Let the vectors $\mathbf{v}^a$ and $\mathbf{v}^b$ be the configurations of the two systems. Then from $v_i^a=\sum_jw_j^aD_{ji}$ the overlap between replicas $a$ és $b$ in terms of the portfolio weights is
\begin{equation}
Q^{ab}=\frac{1}{N}\sum_{i=1}^Nv_i^av_i^b=\frac{1}{N}\sum_{i=1}^N\sum_{j=1}^N\sigma_{ij}w_i^aw_j^b.
\end{equation}
Therefore, $NQ^{ab}$ is nothing but the true covariance (calculated on the basis of complete information, $r\to 0$) of the portfolios $\mathbf{w}^a$ and $\mathbf{w}^b$ . In particular, $NQ^{aa}$ is the true variance (risk) of the portfolio $\mathbf{w}^a$.

As the portfolio weights have been normalized as $\sum_iw_i=N$~, the standard deviation of the true optimum will be $N\sigma^*(\mu)$, instead of $\sigma^*(\mu)$ . Then the estimation error $q_0$ of the portfolio $\mathbf{w}^a$ will be a simple function of $Q^{aa}$:
\begin{equation}
q_0(\mathbf{w}^a)=\frac{Q^{aa}}{N{\sigma^*}^2(\mu)}.
\end{equation}
The expected error of the estimated optimum is therefore equal to the above expression in the thermodynamic limit, at zero temperature. As the energy surface is strictly convex, we expect that at low temperature every replica tends to the same minimum, therefore $Q^{aa}$ will be independent of the replica index. (In the next subsection this will be explicitly shown to be the case.) Accordingly, 
\begin{equation}
\mathbb{E}\left[q_0\right]=\frac{1}{\nu(\mu)}\lim_{\beta\to\infty}\lim_{N\to\infty}Q^{aa}\label{eq:q0replica}
\end{equation}
independently of $a$. The equilibrium value of $Q^{aa}$ will be calculated in the next subsection.

\subsection{The saddle point}

As we wish to calculate the integral \eqref{eq:beforesp} in the limit $N\to\infty$ , we can use the saddle point method:

\begin{equation}
\int\prod_{ab}dQ^{ab}\int\prod_{ab}d\hat{Q}^{ab}e^{-N\left[G(\mathbf{Q},\hat{\mathbf{Q}};\beta)+O(1/N)\right]}\sim e^{-N\min_{\mathbf{Q},\hat{\mathbf{Q}}}G(\mathbf{Q},\hat{\mathbf{Q}};\beta)},
\end{equation}
which becomes exact in the thermodynamic limit. The saddle point conditions are:

\begin{equation}
2\frac{\partial G_{\beta}}{\partial Q^{ab}}=-\widehat{Q}^{ab}+\beta\left[\left(r\beta\mathbf{Q}+\mathbf{I}\right)^{-1}\right]^{ba}=0,\label{eq:spQ}
\end{equation}
\begin{equation}
2\frac{\partial G_{\beta}}{\partial\widehat{Q}^{ab}}=-Q^{ab}+\nu(\mu)+\left[\widehat{\mathbf{Q}}^{-1}\right]^{ba}=0.\label{eq:spQhat}
\end{equation}

The solution is not hard and leads us to the following result:
\begin{eqnarray}
Q_{sp}^{ab}&=&\frac{1}{1-r}\left[\nu(\mu)+\beta^{-1}\delta^{ab}\right],\\
\hat{Q}_{sp}^{ab}&=&\beta(1-r)\left[\frac{\beta r\nu(\mu)}{\beta r\nu(\mu)+1}+\delta^{ab}\right],
\end{eqnarray}
where the subscript $sp$ signifies the saddle point. We can see therefore that at the saddle point the overlap matrix and its conjugate are invariant w.r.t. the permutation of replicas, the only thing that matters is whether we are considering the same replica or distinct ones. In other words, the saddle point is replica symmetric. As we mentioned in the previous section, this is a consequence of the fact that, independently of the sample, the Hamiltonian and the constraints determine a single, unique ground state, provided $r<1$, that is the sample size is sufficiently large. 

It can be seen furthermore that the difference between the diagonal and off-diagonal elements of $\mathbf{Q}_{sp}$ is proportional to $\beta^{-1}$, so it vanishes at zero temperature, which is explained by the fact that all the replicas settle into the same equilibrium state, so the difference between the self-overlap and the overlap of different replicas will become zero. 

At this point, from \eqref{eq:q0replica} we can explicitly determine $\mathbb{E}\left[q_0\right]$:
\begin{equation}
\mathbb{E}\left[q_0\right]=\lim_{\beta\to\infty}\frac{1}{1-r}\left(1+\frac{1}{\beta\nu(\mu)}\right)=\frac{1}{1-r}.\label{eq:q0markowitz}
\end{equation}
This result is in agreement with the results obtained in the special case of the global minimum and i.i.d. zero mean returns in \cite{pafka2003Noisy} and \cite{Burda2003Econophysics}. The most surprising feature of \eqref{eq:q0markowitz} is that for large portfolios the average estimation error depends only on the ratio $N/T$, completely independently of the parameters $\sigma_{ij}$ and $\mu_i$ of the distribution, and also of the expected return $\mu$.

In order to obtain the free energy density, let us substitute \eqref{eq:spQ} and \eqref{eq:spQhat} into \eqref{eq:exponent}. Writing $Q^{ab}=q+\Delta q\delta^{ab}$ we have $\mathrm{Tr}\ln\mathbf{Q}=n\left(q/\Delta q+\ln\Delta q\right)+O(n^2)$, and after some algebra we find
\begin{equation}
G(\mathbf{Q}_{sp},\hat{\mathbf{Q}}_{sp};\beta)=\frac{1}{2}n\beta\nu(\mu)(1-r)+O(n^2).
\end{equation}
According to Subsection \ref{sec:replicatrick} we can then write the free energy density as follows:
\begin{equation}
\mathbb{E}[f(\beta,r)]=\frac{f_0}{\beta}+\lim_{n\to0}\frac{1}{n\beta}G(\mathbf{Q}_{sp},\hat{\mathbf{Q}}_{sp};\beta)=\frac{f_0}{\beta}+\frac{1}{2}\nu(\mu)(1-r),
\end{equation}
where the additive term $f_0$ comes from the multiplicative constants in $\mathbb{E}[Z^n]$. Reviewing the integrals we can see that $f_0$ does not depend on $\beta$, and because of the limits $n\to0$ and $N\to\infty$ it does not depend on $n$, $N$ and $T$ either.
 
The zero temperature limit of the free energy is the minimum value of the Hamiltonian, so for large $N$ the estimated risk at the optimum, averaged over the samples, is:
\begin{equation}
\mathbb{E}\left[{\hat\sigma}^{*2}(\mu)\right]=2N\lim_{\beta\to\infty}\mathbb{E}[f(\beta,r)]=N^2{\sigma^*}^2(\mu)(1-r).
\end{equation}

The estimated in-sample loss is thus seen to vanish inversely proportionally to the out of sample estimation error at the critical point for a generic covariance matrix and all along the efficient frontier. This phenomenon was first observed in numerical experiments and confirmed analytically for a diagonal covariance matrix at the global minimum of the variance in \cite{pafka2003Noisy}. The same inverse proportionality was found in the case of the ES risk measure \cite{kondor2015Contour}. These findings demonstrate how grossly in-sample estimates can underestimate risk.

We conclude this subsection with and important remark. The fact that we could determine the extremum of \eqref{eq:exponent}, was a very lucky happenstance, due to the simplicity of the objective function. In more complicated cases the usual procedure is to assume replica symmetry, express the  $G(\mathbf{Q},\hat{\mathbf{Q}};\beta)$ as a function of the diagonal and off-diagonal elements of $\mathbf{Q}$ and $\hat{\mathbf{Q}}$, and minimize over these variables. As the saddle point is, in general, not necessarily replica symmetric, one has to investigate the stability of the replica symmetric saddle point. In some cases (for example, in spin glasses) replica symmetry breaks down at a non-zero temperature, and in the region below this a symmetry breaking solution becomes stable \cite{mezard1987Spin}. Our present problem is much simpler, and the stability of the replica-symmetric saddle point is preserved all the way down to zero temperature. This is demonstrated in the next subsection.

\subsection{The stability of the replica-symmetric saddle point}

The saddle point is stable, if the functional $G(\mathbf{Q},\hat{\mathbf{Q}};\beta)$ takes its minimum at the point $\mathbf{Q}_{sp}$. Since the conjugate variable is complex and the contour of integration, originally running along the imaginary axis, can be deformed, it does not matter whether the extremum of $G$ with respect to $\hat{\mathbf{Q}}$ is a minimum or maximum. Therefore, it is sufficient to check the stability of $G(\mathbf{Q};\beta)=G(\mathbf{Q},\hat{\mathbf{Q}}_{sp};\beta)$ at $\mathbf{Q}_{sp}$. Expressing $\hat{\mathbf{Q}}$ from \eqref{eq:spQhat} and substituting into \eqref{eq:exponent} we obtain
\begin{equation}
G(\mathbf{Q};\beta)=\frac{1}{2}\left[-n-\mathrm{Tr}\ln\left(\mathbf{Q}-\nu(\mu)\mathbf{U}\right)+\frac{1}{r}\mathrm{Tr}\ln\left(\beta r\mathbf{Q}+\mathbf{I}\right)\right],
\end{equation}
where $\mathbf{U}$ denotes the $n\times n$ matrix with all its elements equal to 1. Let us now calculate the Hess matrix of $G(\mathbf{Q};\beta)$ (absorbing a factor 2 for simplicity):
\begin{multline}
H^{ab,cd}=2\frac{\partial^2}{\partial Q^{ab}\partial Q^{cd}}G(\mathbf{Q};\beta)= \left[\left(\mathbf{Q}-\nu(\mu)\mathbf{U}\right)^{-1}\right]^{ac}\left[\left(\mathbf{Q}-\nu(\mu)\mathbf{U}\right)^{-1}\right]^{bd}-\\
-r\beta^2\left[\left(\beta r\mathbf{Q}+\mathbf{I}\right)^{-1}\right]^{ac}\left[\left(\beta r\mathbf{Q}+\mathbf{I}\right)^{-1}\right]^{bd}.
\end{multline}
Substituting $\mathbf{Q}_{sp}$ into the Hess matrix, we get $H^{ab,cd}_{sp}=R^{ac}R^{bd}$, where
\begin{equation}
R^{ab}=\rho+\Delta\rho\delta^{ab}=\beta(1-r)^{3/2}\left(-\frac{\beta r \nu(\mu)}{\beta r \nu(\mu)n+1}+\delta^{ab}\right).
\end{equation}
The condition for the stability of the saddle point is that $H^{ab,cd}_{sp}$ be strictly positive definite. In order to check this, let us solve the eigenvalue problem 
\begin{gather}
\sum_{c=1}^{n}\sum_{d=1}^{n}H^{ab,cd}_{sp}S^{cd}_m=\lambda_m S^{ab}_m,
\end{gather} 
where $\lambda_m$ are the eigenvalues, and the number of the eigenvectors $S^{ab}_m$ is $n^2$. We must find that all the eigenvalues are positive. Because of the symmetry of $R^{ab}$ the matrix $H^{ab,cd}_{sp}$ is also symmetric, i.e. $H^{ab,cd}_{sp}=H^{cd,ab}_{sp}$, thus an orthonormed basis can be selected from the eigenvectors $S^{ab}_m$. In addition to this, the Hessian also displays an important further symmetry, namely $H^{ab,cd}_{sp}=H^{ba,dc}_{sp}$. An immediate consequence of this is that whenever $S^{ab}$ is an eigenvector, its transposed $S^{ba}$, its symmetric part $(S^{ab}+S^{ba})/2$ and antisymmetric part $(S^{ab}-S^{ba})/2$ are also eigenvectors, and they belong to the same eigenvalue. The $n(n+1)/2$ dimensional space of symmetric matrices, and the $n(n-1)/2$ dimensional space of antisymmetric matrices are thus invariant subspaces of the Hessian. According to elementary considerations, the invariant subspaces of $H^{ab,cd}_{sp}$ are as shown in the Table below:
\begin{center}
\begin{scriptsize}
\begin{tabular}{|c|c|p{57mm}|c|}
\hline
Subspace & Eigenvalue & \centering Eigenvectors & Multiplicity\\
\hline
I & $\lambda=(\rho n+\Delta\rho)^2$ & \centering $S^{ab}=1$ for all $a$ and $b$ & 1\\
\hline
II & $\lambda=(\rho n+\Delta\rho)\Delta\rho$ & \centering $S^{ab}=\left\{
					\begin{array}{ll}
					1-n & \mbox{if $a=b=k$,}\\
					\frac{2-n}{2} & \mbox{if either $a=k$, or $b=k$,}\\
					1 & \mbox{if $a\ne k$ and $b\ne k$.}\\
					\end{array}
				\right.$ & $n-1$\\
& & \centering  $k=1,2,...,n-1$ & \\
\hline
III & $\lambda=\Delta\rho^2$ & \centering Those symmetric matrices, which leave the vector $(1,1,1,...,1)$ invariant. & $\frac{1}{2}n(n-1)$\\
\hline
IV & $\lambda=(\rho n+\Delta\rho)\Delta\rho$ & \centering $S^{ab}=\delta^{ak}-\delta^{bk}$ & $n-1$\\
& & \centering $k=1,2,...,n-1$ & \\
\hline
V & $\lambda=\Delta\rho^2$ & \centering Those matrices $S^{ab}$ , for which $\sum_{b=1}^nS^{ab}=0$ for arbitrary $a$. & $\frac{1}{2}(n-1)(n-2)$\\
\hline
\end{tabular}
\end{scriptsize}
\end{center}

Symmetric matrices belong to subspaces I, II and III, and by adding up the multiplicities we can see that these subspaces span the full $n(n+1)/2$ dimensional linear space of $n\times n$ symmetric matrices. Similarly, one can convince herself that the invariant subspaces IV and V span the full $n(n-1)/2$ dimensional space of the $n\times n$ antisymmetric matrices. The grand total of multiplicities is $n^2$, so we have found all the eigenvalues and eigenvectors.

Therefore, the Hessian $H^{ab,cd}_{sp}$ has three different eigenvalues:
\begin{align}
\lambda_1=(\rho n+\Delta\rho)^2&=\frac{\beta^2(1-r)^3}{\left(\beta r\nu(\mu)n+1\right)^2},\\
\lambda_2=(\rho n+\Delta\rho)\Delta\rho&=\frac{\beta^2(1-r)^3}{\beta r\nu(\mu)n+1},\\
\lambda_3=\Delta\rho^2&=\beta^2(1-r)^3.
\end{align}
Of these, $\lambda_1$ has multiplicity 1, $\lambda_2$ is $2(n-1)$-fold degenerate, while $\lambda_3$ is $(n-1)^2$-fold degenerate. The eigenvalues are positive as long as $r<1$. Therefore the replica symmetric saddle point is stable at any temperature, provided $r<1$, that is $T>N$. At the critical point $r=1$ all three eigenvalues vanish, the model becomes unstable against fluctuations in any directions. With this we have given the characterisation of the noise sensitivity of the Markowitz problem in the thermodynamic limit.

\section{Summary}
Several years after the first publications on the statistical physics approach to the optimization of the Expected Shortfall \cite{ Ciliberti2007On} and Mean Absolute Deviation \cite{ciliberti2007Risk} risk measures, this paper addressed the problem of optimizing the variance. In contrast to previous studies, which focused on the global minimum, in the present paper, the simplicity of the objective function allowed us to extend the replica method to the full Markowitz problem, by considering a generic covariance matrix and, in addition to the budget constraint, also the constraint on the expected return of the portfolio. From the point of view of the method itself, the most interesting feature we encountered was that the replica symmetric solution emerged as the unique solution, without having to assume this symmetry in advance. It was also straightforward to check the stability of this solution, with a foregone conclusion. These features point in the direction of a possible rigorous foundation of the replica method in the case of convex risk measures. 

The solution we found displayed a remarkable universality: the estimation error turned out to be the same and blowing up as $1/(1-r)$ all along the efficient frontier, independently of the expected return and the details of the covariance matrix. Note that this universality goes far beyond what is usually meant by universality in the theory of critical phenomena. It is not only the exponent of the divergence that is independent of the ``microscopic'' details, but also the critical point itself. This is analogous with the universality discovered by Donoho and Tanner \cite{Donoho2009Observed} in a class of random geometric problems. On the basis of this analogy we conjecture that the critical point of the variance will remain at $r=1$ even if the returns are drawn from a distribution different from the Gaussian considered here. Experience with the minimax risk measure, where the critical point is invariant with respect to the returns distribution (as long as it is symmetric)                        \cite{ kondor2007Noise}, and with the numerical experiments on the phase boundary of ES \cite{caccioli2015Portfolio}, support this conjecture. 
As for the in-sample estimate of the optimal value of the variance, it was found to be $(1-r)$ times its true value. The inverse proportionality of the estimated variance and the out of sample estimation error near the critical point seems to be a recurrent feature of these type of problems: it has also been found in the optimization of ES \cite{caccioli2015Portfolio}, and of the minimax again. 
With the replica approach to variance optimization established, we can turn to finding a cure for the divergent estimation error by regularization. This is left to a subsequent publication.

\section*{Acknowledgement} I.K. thanks Risi Kondor, B\'alint Komj\'ati and Christoph Memmel for helpful discussions.  F.C.  acknowledges support of the Economic and Social Research Council (ESRC) in funding the Systemic Risk Centre (ES/K002309/1).

\bibliographystyle{unsrt}
\bibliography{References1}
\end{document}